\documentstyle[prb,aps,epsf]{revtex}
\draft

\begin{document}

\title{ Spin 1/2 Magnetic Impurity in a 2D Magnetic System Close to\\
Quantum Critical Point.}
\author{O.P. Sushkov}

\address{School of Physics, University of New South Wales,\\
 Sydney 2052, Australia}

\maketitle

\begin{abstract}
We consider a magnetic impurity in a spin liquid state of a magnetic system 
which is close to the quantum phase transition to the magnetically ordered 
state. There is similarity between this problem and the Kondo problem.
We derive the impurity Green's function, consider renormalizations of the 
magnetic moments of the impurity, calculate critical indexes for the magnetic 
susceptibilities and finally consider specific heat and magnetic interaction 
of two impurities.
\end{abstract}


\section{Introduction}
There is no need to explain the importance of the Kondo problem for condense
matter physics. It is relevant to the magnetic impurities in metals,
heavy fermion compounds \cite{Lee}, tunneling phenomena in quantum 
dots \cite{dot}, correlated lattice fermion systems \cite{Zhu}, and
many other physical systems. 
The problem of magnetic impurity in a two-dimensional (2D) insulating system 
with long range antiferromagnetic order has also attracted great interest.
This includes an impurity spin with an on-site \cite{Nagaosa,Igarashi}
and sublattice symmetric \cite{Oitmaa1} coupling, as well as an isolated
ferromagnetic bond \cite{KLee}.  
Because of Adler's relation for the impurity-spin-wave interaction 
these systems have no nontrivial infrared dynamics at zero temperature
in spite of the gapless  spectrum of Goldstone spin waves  \cite{Kotov1}.
This makes the impurity problem for the insulating state much simpler than the 
Kondo  one. However it has been realized recently \cite{hole} that in the case
when the 2D magnetic insulating system is close to the quantum critical point
the infrared dynamics of the impurity is highly nontrivial and
to a large extend is similar to that for the Kondo problem.
The quantum critical impurity problem has been very recently addressed
by Vojta, Buragohain, and Sachdev \cite{Vojta}. They demonstrated 
that there are nontrivial critical indexes for the impurity Green's function
and for  magnetic susceptibilities. 
In the present work we consider the same quantities calculating more
accurately the critical indexes and in some cases the prefactors.
In addition we consider specific heat and interaction of the impurities
\cite{note}.

\section{The Hamiltonian}
To be specific we consider a magnetic impurity in two coupled Heisenberg 
planes.
The two coupled Heisenberg planes is an ideal example of a 2D critical system 
which can be described by O(3) nonlinear $\sigma$-model.
The two planes model is very well studied at zero temperature both numerically 
\cite{Hida,Sandvik,Jaklic,Weihong} and analytically \cite{Morr,Kotov}.
Finite temperature properties are also well 
understood \cite{Sandvik,Oitmaa,Elstner,SSS}.
Hamiltonian of the system with impurity is of the form
\begin{eqnarray}
\label{H1}
&&H=H_2+H_{imp},\\
&&H_2=J\sum_{<ij>}\left({\bf S}_{i}^{(1)}{\bf S}_{j}^{(1)}+{\bf S}_{i}^{(2)}
{\bf S}_{j}^{(2)}\right)
+J_{\perp}\sum_{i}{\bf S}_{i}^{(1)}{\bf S}_{i}^{(2)},\nonumber\\
&&H_{imp}=j{\bf S}_{0}^{(1)}{\bf s}.\nonumber
\end{eqnarray}
Here ${\bf S_i^{(n)}}$ is spin 1/2 on the square lattice. the index $i$ numerates 
cites, and
the index $n$ numerates planes. $J$ is an antiferromagnetic coupling in the plane, and
$J_{\perp}$ is an antiferromagnetic coupling between the planes. Spin of the impurity,
$s=1/2$, is coupled to one of the planes. It is known that in this system at 
$J_{\perp}=2.525\pm 0.002J$ \cite{SSS} there is
a quantum phase transition from quantum disordered state to the Neel
state. In the present work we consider only the quantum disordered phase including the
critical point. As we have already pointed out an interesting regime arises only 
close to the critical point and therefore we concentrate on the vicinity of this point. 
Note that in terms of the non-linear $O(3)$ $\sigma$-model an effective Lagrangian of the 
system is 
$L=\partial_{\mu}{\vec \varphi}\partial_{\mu}{\vec \varphi}-m^2{\vec \varphi}^2
+\gamma {\vec \varphi}{\vec s}$, ${\vec \varphi}^2=a^2$,
where the parameters can be expressed in terms of parameters of the original 
Hamiltonian (\ref{H1}).
In our considerations we will use only the original Hamiltonian (\ref{H1}).

Using bond operator representation \cite{BR}
\begin{equation}
\label{t}
{\bf S}^{(1,2)}={1\over 2} \left(\pm {\bf t} \pm {\bf t^{\dag}}-i {\bf t^{\dag}}\times
 {\bf t}\right),
\end{equation}
the two plane Hamiltonian $H_2$ from (\ref{H1}) can be rewritten in terms of the operators
${\bf t_i}=(t_{i,x},t_{i,y},t_{i,z})$, and then  diagonalized by a combination of the  
Fourier and Bogoliubov
transformations with account of the hard core constraint, see Refs. \cite{Kotov,SSS}:
\begin{eqnarray}
\label{huv}
&&{\bf t}_i=\sum_{\bf q}e^{i{\bf q r_i}}(u_{\bf q}t_{\bf q}+v_{\bf q}{\bf t}^{\dag}_{\bf q}),\\
&&H_2 \to \sum_{\bf q}\omega_{\bf q}{\bf t}^{\dag}_{\bf q}{\bf t}_{\bf q},\nonumber
\end{eqnarray}
where 
$u_{\bf q},v_{\bf q}=\sqrt{{{A_{\bf q}}\over{2\omega_{\bf q}}}\pm {1\over 2}}$ are
Bogoliubov parameters.
The operator ${\bf t}_{\bf q}^{\dag}$ creates quasiparticle of the system. This
quasiparticle has spin 1 and we call it spin wave or magnon.
Near the critical point the spin-wave excitation energy is 
\begin{equation}
\label{oo}
\omega_{\bf q}\approx \sqrt{\Delta^2+c^2q^2},
\end{equation}
where $\Delta \ll J$ is the spin-wave gap and $c\approx 1.9 J$ is the spin-wave velocity.
The function $A_{\bf q}$ in the vicinity of the critical point is q-independent:
$A_{\bf q}\approx A \approx 2.4J$.

In the bond operator representation the impurity Hamiltonian is of the form
\begin{equation}
\label{Himp}
H_{imp}={j\over 2} {\bf s} \left({\bf t_0}+{\bf t_0}^{\dag}-i {\bf t_0}^{\dag}\times 
{\bf t_0}\right) \to {j\over 2} {\bf s} \left({\bf t_0}+{\bf t_0}^{\dag}\right).
\end{equation}
Here we have dropped the term ${\bf t_0}^{\dag}\times {\bf t_0}$. The matter is that we are 
interested in nontrivial long-range dynamics, but one can prove that all diagrams 
generated by the ${\bf t_0}^{\dag}\times {\bf t_0}$ term are infrared convergent, 
and therefore its contribution is less important. An alternative way is to redefine 
$H_{imp}$ as $H_{imp} \to {j\over 2} \ 
{\bf s}\left({\bf S_0}^{(1)}-{\bf S_0}^{(2)}\right)$, 
then the ${\bf t_0}^{\dag}\times {\bf t_0}$ term is canceled out exactly. Note that 
for an integer impurity spin the ${\bf t_0}^{\dag}\times {\bf t_0}$ term could be much 
more important because it can give a bound state of the spin wave with the impurity, 
and hence can lead to the full screening of the impurity.

Using (\ref{huv}) we rewrite $H_{imp}$ in terms of quasiparticle operators ${\bf t}_{\bf q}$.
\begin{equation}
\label{Himp1}
H_{imp}={j\over 4}\sum_{\bf q}(u_{\bf q}+v_{\bf q}){\vec t}_{\bf q}^{\dag}{\vec \sigma} + h. c.
\approx{{j\sqrt{A}}\over{2\sqrt{2}}}\sum_{\bf q}{1\over{\sqrt{\omega_{\bf q}}}}
{\vec t}_{\bf q}^{\dag}{\vec \sigma} + h. c.,
\end{equation}
where ${\vec \sigma}$ is the impurity Pauli matrix.
Hereafter we set $J=1$, so all energies are measured in units of $J$.

\section{The impurity Green's function at zero temperature}

Let us calculate the impurity self energy $\Sigma$ shown in Fig. 1.
\begin{figure}[h]
\vspace{-2pt}
\hspace{-35pt}
\epsfxsize=7cm
\centering\leavevmode\epsfbox{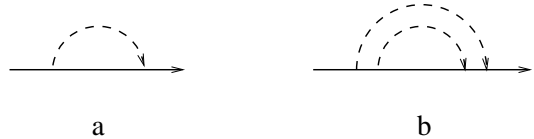}
\vspace{8pt}
\caption{\it {a) Single loop impurity self energy. 
b) Two loop noncrossing contribution to the self energy.
The solid line corresponds to the impurity and the dashed line
corresponds to the spin wave.}}
\label{Fig1}
\end{figure}
\noindent
In single loop approximation,
see Fig.1a, using eq. (\ref{Himp1}) we find
\begin{equation}
\label{s1}
\Sigma^{(1)}(\epsilon)={{3Aj^2}\over{8}}\int{1\over{\omega_{\bf q}(\epsilon-\omega_{\bf q})}}
{{d^2 q}\over{(2\pi)^2}}.
\end{equation}
Close to the critical point $\omega_{\bf q}=\sqrt{c^2 q^2+\Delta^2}$, and
hence simple  integration gives for $\epsilon =0$ (i.e. at the position of the
 quasiparticle pole)
\begin{equation}
\label{s11}
\Sigma^{(1)}(0)=-\alpha^2 \ln{{\Lambda}\over{\Delta}}.
\end{equation}
Here
\begin{equation}
\label{al}
\alpha^2={{3Aj^2}\over{16\pi c^2}}\approx 0.04 j^2,
\end{equation}
is dimensionless coupling constant,
and $\Lambda \sim 2 J$ is the ultraviolet cutoff. Similar calculation for
the second order self energy shown in Fig.1b gives
\begin{equation}
\label{s2}
\Sigma^{(2)}(0)=-{{\alpha^4}\over{\Delta}} \ln{{\Lambda}\over{\Delta}}.
\end{equation}
As soon as $\alpha^2/\Delta \ll 1$ the second order self energy is small
compared to the first order one,
$\Sigma^{(2)} \ll \Sigma^{(1)}$, and hence the perturbation theory is
justified. However at $\alpha^2/\Delta > 1$ the expansion does not
converge and hence one has to sum all orders of perturbation theory.
Exactly at the critical point, $\Delta=0$, the expansion diverges at
arbitrary small $\alpha$.

The problem under consideration has a small parameter which is independent of
the interaction. This is 1/N, where N=3 is number of
components of the spin-wave excitation (O(N) $\sigma$-model).
In the leading in N approximation only the rainbow diagrams
contribute to the impurity self energy. Summation of these diagrams leads to the 
usual noncrossing approximation
(= self consistent Born approximation) for the impurity Green's function
\begin{equation}
\label{Dy}
G(\epsilon)={{1}\over{\epsilon-\Sigma(\epsilon)}},
\end{equation}
\begin{equation}
\label{SB}
\Sigma(\epsilon)=\alpha^2\int G(\epsilon-\omega)d\omega.
\end{equation}
It is convenient to rescale the variables:
$\epsilon\to \epsilon/\alpha^2$,
$\omega \to \omega/\alpha^2$,
$G \to \alpha^2 G$,
$\Sigma \to \Sigma/\alpha^2$.
In the new variables $\alpha$ disappears from the eqs. (\ref{Dy},\ref{SB}). Dependence
on $\alpha$ remains only in the limits of $\omega$-integration: 
$\Sigma=\int_{\Delta_{\alpha}}^{\Lambda_{\alpha}}G d\omega$, where $\Delta_{\alpha}=
\Delta/\alpha^2$ and $\Lambda_{\alpha}=\Lambda/\alpha^2$.
Consider first the critical point, i.e. $\Delta_{\alpha}=0.$
In this case eqs. (\ref{Dy},\ref{SB}) can be solved analytically. The answer is
\begin{equation}
\label{g1}
G(\epsilon)=-{1\over{\sqrt{2(\epsilon_0-\epsilon)}}},
\end{equation}
where $\epsilon_0$ is the impurity binding energy. Eq. (\ref{g1}) is valid if
$|\epsilon_0-\epsilon | \lesssim 1$. For illustration we present in Fig. 2
spectral function, $-{1\over{\pi}}\mbox{Im} G(\epsilon)$, obtained by direct 
numerical solution of eqs. (\ref{Dy},\ref{SB}) at $\Delta_{\alpha}=0$ and 
$\Lambda_{\alpha}=1,2,5$.
\begin{figure}[h]
\vspace{-20pt}
\hspace{-35pt}
\epsfxsize=11cm
\centering\leavevmode\epsfbox{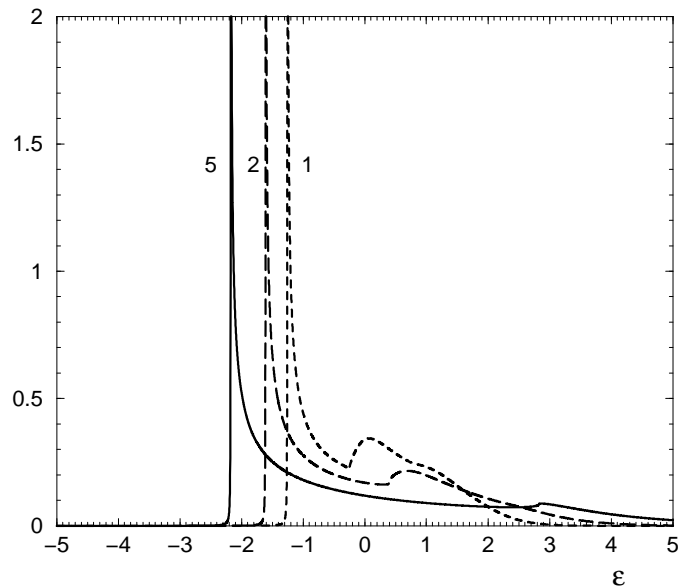}
\vspace{0pt}
\caption{\it {The impurity spectral function 
$-{1\over{\pi}}\mbox{Im} G(\epsilon)$ at the critical point.
The curves obtained in the noncrossing approximation for different values
of the ultraviolet cutoff:  $\Lambda_{\alpha}=1,2,5$.}}
\label{Fig2}
\end{figure}
\noindent
Agreement with analytical solution (\ref{g1}) is perfect. At the large
$\Lambda_{\alpha}=\Lambda/\alpha^2$ the impurity binding energy 
$\epsilon_0 \approx -2.5$. In the original variables it means that
$\epsilon_0 \approx -2.5\alpha^2$, when $\alpha^2 \ll 1$.

As one shall expect the Green's function (\ref{g1}) has no quasiparticle pole.
However the pole appears away from the critical point when the spin-wave gap 
is nonzero .  For illustration we
present in Fig. 3 the spectral functions obtained by numerical solution of
Eqs. (\ref{Dy},\ref{SB}) for $\Lambda_{\alpha}=2$, $\Delta_{\alpha}=0$, and
$\Lambda_{\alpha}=2$, $\Delta_{\alpha}=0.05$.
\begin{figure}[h]
\vspace{-20pt}
\hspace{-35pt}
\epsfxsize=11cm
\centering\leavevmode\epsfbox{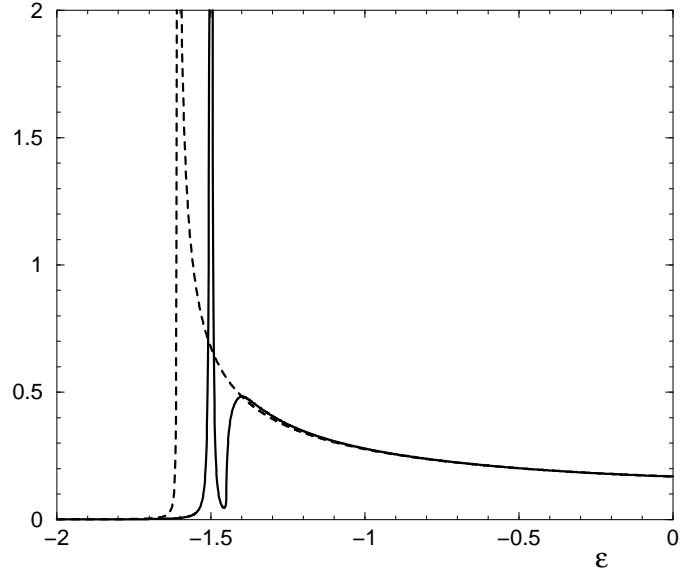}
\vspace{0pt}
\caption{\it {The impurity spectral function 
$-{1\over{\pi}}\mbox{Im} G(\epsilon)$ obtained in the noncrossing approximation.
The dashed curve corresponds to the critical point ($\Delta_{\alpha}=0$),
and the solid curve corresponds to the spin-wave gap 
$\Delta_{\alpha}=0.05$. In both cases the ultraviolet cutoff  is
$\Lambda_{\alpha}=2$.}}
\label{Fig3}
\end{figure}
\noindent
It is clear that the quasiparticle peak absorbs spectral weight of the
incoherent Green's function (\ref{g1}) from the area $\epsilon_0-\epsilon
\lesssim \Delta_{\alpha}$. Therefore the quasiparticle residue 
$Z \propto \sqrt{\Delta_{\alpha}}$  (c.f. with Ref. \cite{hole}). Slightly more
detail analysis of the Eqs. (\ref{Dy},\ref{SB}) shows that
\begin{equation}
\label{zz}
Z \approx 0.8 \sqrt{\Delta_{\alpha}}.
\end{equation}

We have considered above the leading in N approximation. Let us
estimate now $1/N$ correction which is due to the single loop contribution to the
impurity-magnon vertex function shown in Fig.4. 
\begin{figure}[h]
\vspace{-2pt}
\hspace{-35pt}
\epsfxsize=11cm
\centering\leavevmode\epsfbox{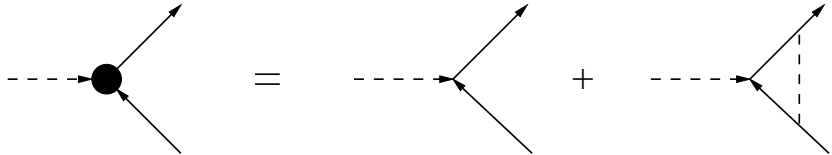}
\vspace{8pt}
\caption{\it {The impurity - spin wave vertex with account of
single loop correction.
The solid line corresponds to the impurity and the dashed line
corresponds to the spin wave.}}
\label{Fig4}
\end{figure}
\noindent
Taking into account the
algebraic relation for the Pauli matrices,
$\sigma_{\mu}\sigma_{\nu}\sigma_{\mu}=-\sigma_{\nu}$,
we find the vertex function given by Fig. 4
\begin{equation}
\label{v1}
\Gamma (\epsilon,\lambda) = \Gamma_{bare} \left(1-{1\over N}
\int G(\epsilon-\omega) G(\epsilon-\omega-\lambda)d\omega \right).
\end{equation}
Here $\epsilon$ and $\lambda$ are energies of incoming impurity and spin wave
correspondingly, and $\Gamma_{bare}$ is the bare vertex given by eq. (\ref{Himp1}).
Taking $\lambda \sim \epsilon-\epsilon_0$ and using Green's function (\ref{g1})
we find after integration in (\ref{v1})
\begin{equation}
\label{v2}
\Gamma_{\lambda} = \Gamma_{bare} \left(1-{1\over{2 N}}\ln {{\Lambda_{\alpha}}\over{\lambda}}\right).
\end{equation}
This is the the first term of $1/N$ expansion, and keeping in mind scaling
behavior we find 
\begin{equation}
\label{v3}
\Gamma_{\lambda} \propto \lambda^{y}, \ \ \ \ y \approx{1\over {2N}}={1\over 6}.
\end{equation}

The impurity self energy is given by the diagram presented in
Fig. 5. 
\begin{figure}[h]
\vspace{-2pt}
\hspace{-35pt}
\epsfxsize=5cm
\centering\leavevmode\epsfbox{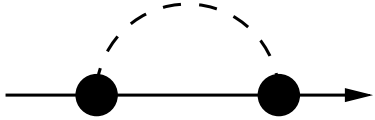}
\vspace{8pt}
\caption{\it {The impurity self energy with account of vertex
corrections.}}
\label{Fig5}
\end{figure}
\noindent
It differs from (\ref{SB}) by the vertexes: the bare  vertexes are replaced by
the ``exact'' ones given by eq. (\ref{v3}). Effectively this introduces an additional factor
$\omega^{2y}$ in the integrand in eq. (\ref{SB}). Assuming power behavior of the
Green's function
\begin{equation}
\label{g2}
G(\epsilon) \propto {1\over{(\epsilon_0-\epsilon)^x}}
\end{equation}
and performing integration we find
\begin{equation}
\label{s3}
\Sigma(\epsilon) \propto (\epsilon_0-\epsilon)^{2y-x+1} + const.
\end{equation}
Substitution of this self energy into Dyson equation (\ref{Dy}) gives the following
condition of self consistency
\begin{equation}
\label{Dy1}
(\epsilon_0-\epsilon)^{-x} \propto (\epsilon_0-\epsilon)^{x-2y-1}.
\end{equation}
Therefore the Green's function critical index with account of the leading $1/N$
correction is
\begin{equation}
\label{ci}
x={1\over 2}+y={1\over 2}(1+{1\over{N}}) \approx 0.67
\end{equation}

Away from the critical point the quasiparticle pole appears in the Greens
function. It absorbs spectral weight of the incoherent Green's function
(\ref{g2})  from the area $\epsilon_0-\epsilon \lesssim \Delta_{\alpha}$. 
Therefore the quasiparticle residue scales as
\begin{equation}
\label{zz1}
Z \propto \Delta_{\alpha}^z, \ \ \ z=1-x = {1\over 2}-y=
{1\over 2}(1-{1\over{N}}) \approx 0.33.
\end{equation}

\section{Magnetic moments of the impurity, Susceptibilities}
Following  Vojta, Buragohain, and  Sachdev \cite{Vojta} 
we consider two different types of the magnetic interaction. 
The first one is an interaction when the magnetic field $h$ interacts only with the 
impurity
\begin{equation}
\label{HM1}
\tilde H_M^{(I)}=-2{\bf s}{\bf h}.
\end{equation}
The second case is homogeneous magnetic field
\begin{equation}
\label{HM2}
H_M^{(II)}=-2 {\bf s}{\bf h} -\sum_{i,n}2{\bf S_i^{(n)}}{\bf h}.
\end{equation}

It is clear that in the first case the renormalized magnetic moment is
proportional to the quasiparticle residue $Z$, and hence, according to
(\ref{zz1}) it scales as
\begin{equation}
\label{m1}
\mu^{(I)} \propto \Delta^z.
\end{equation}
If we consider the system exactly at the critical point, but at finite
temperature, then the effective spin-wave gap is equal to the temperature, see e.g.
Refs. \cite{CSY,SSS}:
\begin{equation}
\label{gt}
\Delta \to \Delta_T \approx 0.962 T.
\end{equation}
Together with (\ref{m1}) this gives the following dependence of the impurity
magnetic susceptibility on temperature.
\begin{equation}
\label{xi1}
\chi_{imp}^{(I)}={{\mu^2}\over{T}} \propto {1\over{T^{2y}}}\approx {1\over{T^{1/N}}}= 
{1\over{T^{0.33}}}.
\end{equation}

For homogeneous magnetic field  the impurity magnetic moment is not
renormalized because the interaction (\ref{HM2}) is proportional to the total
spin which is conserved. The magnetic moment is certainly redistributed over
the volume of size $r\sim 1/\Delta_T$, but the value is conserved.
For this reason the impurity susceptibility is given by usual the Curie law.
\begin{equation}
\label{xi11}
\chi_{imp}^{(II)}={{1}\over{T}}.
\end{equation}
This conclusion is not quite trivial since the magnon cloud size 
$r \to \infty$ at $T \to 0$.
Besides eq. (\ref{xi11}) does not agree with a conclusion from the paper \cite{Vojta}.
Therefore in the next section we also present a diagrammatic
prove of our statement. It is rather technical section and a reader who is satisfied  
by the general arguments can skip it.

\section{Corrections to the impurity magnetic moment}

In the present section only homogeneous magnetic field (\ref{HM2}) is considered,
therefore to simplify notations we omit the superscript (II).
We restrict our consideration by single loop corrections,
and follow the way used in Ref. \cite{SSS} for calculation
of the spin-wave magnetic moment.
First we discuss a zero temperature case. Single loop self energy
is given by eq. (\ref{s11}). This as a
simple perturbation theory in $\alpha$, but one can also consider this as
a contribution to renormalization group equations with running coupling
constant $\alpha$ and running infrared cutoff $\Delta$.
Anyway, correction to the quasiparticle residue due to 
(\ref{s11}) is following
\begin{equation}
\label{dz}
\delta Z = -{{\partial \Sigma^{(1)}}\over{\partial \Delta}}
=-{{\alpha^2}\over{\Delta}}.
\end{equation}
The impurity magnetic moment is renormalized according to the diagrams shown in Fig. 6.
\begin{figure}[h]
\vspace{-2pt}
\hspace{-35pt}
\epsfxsize=10cm
\centering\leavevmode\epsfbox{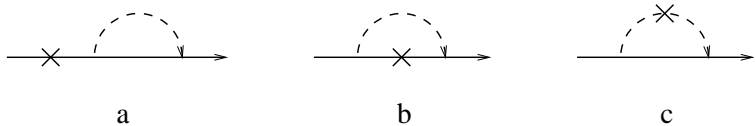}
\vspace{8pt}
\caption{\it {Zero temperature single loop corrections to the impurity magnetic 
moment. The cross denotes an external magnetic field attached to the corresponding
line.}}
\label{Fig6}
\end{figure}
\noindent
First contribution comes from the correction to the quasiparticle residue:
$\delta \mu_{6a}= \delta Z$. Straightforward calculation gives
$\delta \mu_{6b}= -\alpha^2/(3\Delta)$ and
$\delta \mu_{6c}= 4\alpha^2/(3\Delta)$.
Altogether this gives
\begin{equation}
\label{dmu1}
\delta \mu_6= \delta\mu_{6a}+\delta\mu_{6b}+\delta\mu_{6c}=0.
\end{equation} 
So as one shall expect there is no renormalization of the magnetic moment.

At finite temperature the relation (\ref{dmu1}) remains valid since the only
thing we have to do is to replace $\Delta \to \Delta_T$.
However at finite temperature there are also additional diagrams which are
due to the heat bath of the excited magnons \cite{Mats}. First of all these 
are the two contributions to the self energy shown in Fig. 7 
\begin{figure}[h]
\vspace{8pt}
\hspace{-35pt}
\epsfxsize=6cm
\centering\leavevmode\epsfbox{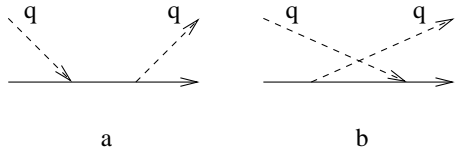}
\vspace{2pt}
\caption{\it {Temperature induced contributions to the impurity self energy.
Dashed lines denote magnons from the heat bath.}}
\label{Fig7}
\end{figure}
\noindent
Both contributions are proportional to the magnon mean occupation number
\begin{equation}
\label{nq}
n_{\bf q}={{1}\over{e^{\omega_{\bf q}/T}-1}},
\end{equation}
but they are of the opposite sign and exactly cancel each other. So they
do not influence the position of the quasiparticle pole. However these
diagrams contribute equally to the quasiparticle residue
\begin{equation}
\label{dz1}
\delta Z_T =-2\alpha^2 \int {{n_{\bf q} d\omega_{\bf q}}\over{\omega_{\bf q}^2}}
=-{{2\alpha^2}\over{\Delta_T}}\int_1^{\infty}{{dx}\over{x^2(e^x-1)}}.
\end{equation}
This gives a correction to the inpurity magnetic moment. Other
thermally  induced corrections to the  magnetic moment are
given by diagrams presented in Fig. 8.
\begin{figure}[h]
\vspace{8pt}
\hspace{-35pt}
\epsfxsize=14cm
\centering\leavevmode\epsfbox{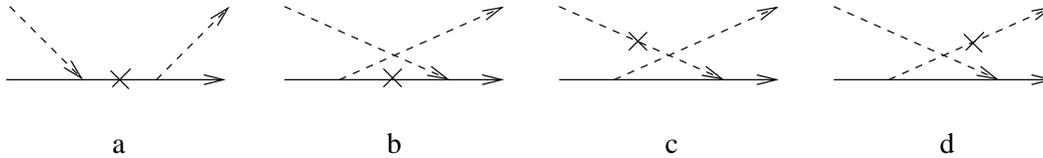}
\vspace{2pt}
\caption{\it {Thermal corrections to the impurity magnetic moment.
Dashed lines denote magnons from the heat bath.
The cross denotes an external magnetic field attached to the corresponding
line.}}
\label{Fig8}
\end{figure}
\noindent
Straightforward calculation gives $\delta\mu_{8a}=\delta\mu_{8b}=
-{{\alpha^2}\over{3\Delta_T}}\int_1^{\infty}dx/[x^2(e^x-1)]$ , and\\
$\delta\mu_{8c}=\delta\mu_{8d}=
-{{4\alpha^2}\over{3\Delta_T}}\int_1^{\infty}dx/[x^2(e^x-1)]$.
Total thermally induced correction to the impurity magnetic moment is equal to
\begin{equation}
\label{dmu2}
\delta \mu_T= \delta Z_T+\delta\mu_{8a}+\delta\mu_{8b}+\delta\mu_{8c}
+\delta\mu_{8d}=0.
\end{equation} 
Together with eq. (\ref{dmu1}) this proves that the impurity magnetic moment
is not renormalized.

Thus, in spite of the cloud of virtual and thermal magnons, effectively
the impurity in the external magnetic field can be described as a spin 1/2
system with unrenormalized magnetic moment. Hence we immediately come
to eq. (\ref{xi11}). We would like to stress that the arguments
related to the magnetic moment guarantee only singular in $T$ part of the
susceptibility, therefore instead of (\ref{xi11}) it is more correct to write
$\chi_{imp}^{(II)}=1/T + const$.

\section{Specific heat related to the impurity}
Binding energy of the impurity has been calculated in section III.
Exactly at the critical point, at zero temperature, and at small
coupling constant, $\alpha^2 \ll 1$, the binding energy is 
$\epsilon_0 \approx -2.5\alpha^2$. A finite spin-wave gap $\Delta$ pushes the
position of the quasiparticle pole up, see Fig. 3. For a small gap the dependence of 
$\epsilon_0$ on the gap  is linear, and the coefficient of the proportionality is 
approximately 2:
\begin{equation}
\label{e00}
\epsilon_0 \approx \alpha^2(-2.5+2\Delta).
\end{equation}
The origin of $\Delta$ in this equation is not important: whether $\Delta$ is nonzero because
the system is away from the critical point, or the system is at the critical point, but
$\Delta$ is nonzero because of temperature. In the later case we can use eq. (\ref{gt})
for the gap and hence the impurity specific heat is
\begin{equation}
\label{cimp}
C_{imp}={{d{\epsilon_0}}\over{dT}} \approx 2\alpha^2.
\end{equation}
Note that this is highly unusual result because $C_{imp}\ne 0$ at $T \to 0$.
For comparison: the bulk specific heat of a 2D antiferromagnet 
at the quantum critical point is quadratic in temperature, $C_{bulk} \propto T^2$,
see e.g. Refs. \cite{CSY,SSS}.
From (\ref{cimp}) one concludes that the impurity entropy is $S_{imp}=\int C_{imp}dT/T
\propto \ln T$. Strictly speaking this is nonsense because it gives $S_{imp}(T=0)=-\infty$.
This probably indicates that there is a small nonzero critical index $\xi$ in the
specific heat dependence: $C_{imp} \propto T^{\xi}$. Equation (\ref{e00}) has been
derived in the noncrossing approximation. One can check that single loop vertex corrections
considered at the end of the section III do not change this equation. It probably means 
that nonzero $\xi$ can appear only due to the two loop corrections, 
$\xi \propto 1/N^2$.
There is no doubt that this is a very interesting problem which deserves further
analytical analysis and which can be also studied in numerical simulations.

\section{Static interaction between two distant impurities}
Interaction between two impurities is given by diagram shown in Fig. 9.
\begin{figure}[h]
\vspace{8pt}
\hspace{-35pt}
\epsfxsize=3cm
\centering\leavevmode\epsfbox{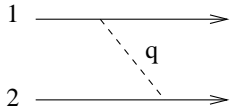}
\vspace{8pt}
\caption{\it {Spin-wave exchange between two impurities.}}
\label{Fig9}
\end{figure}
\noindent
Since the impurities are localized, we have to integrate over all possible
momenta transfer q. Away from the critical point there are two distinct
regimes: $c/r < \Delta$ and $c/r > \Delta$, where $r$ is distance between the
impurities and $c\approx 1.9J$ is the spin-wave velocity. In the first regime
the interaction drops down exponentially with distance and it is not an interesting 
case. We consider only $c/r > \Delta$ case which is also relevant to the critical point.
We also assume that the coupling constant is strong enough: $\alpha^2/\Delta > 1$.
The impurity-magnon interaction is given by the Hamiltonian (\ref{Himp1}).
According to the consideration in section III it is renormalized as 
$H_{imp} \to Z_q \Gamma(q)$,
where $Z_q$ is quasiparticle residue (\ref{zz}), and $\Gamma(q)$ is the vertex
(\ref{v3}).  Therefore the
interaction corresponding to the diagram Fig.9 is of the form
\begin{equation}
\label{vr}
V(r) =
\int{{Z_q^2\Gamma^2_qe^{i{\bf q r}}}\over{\omega_{\bf q}}}{{d^2 q}\over{(2\pi)^2}}=
-{\vec \sigma_1}\cdot{\vec \sigma_1} {{Aj^2}\over{4}} 
\int{{\left[Z_q \Gamma_q/\Gamma_{bare}\right]^2
e^{i{\bf q r}}}\over{\omega_{\bf q}^2}}{{d^2 q}\over{(2\pi)^2}},
\end{equation}
where ${\vec \sigma_1}$ and ${\vec \sigma_2}$ are the impurities Pauli matrixes.
One power of $\omega_{\bf q}$ in the denominator appears because of the spin-wave
propagator and another power is due to the bare interaction (\ref{Himp1}). 
With account of (\ref{al}), (\ref{zz}), and (\ref{v3}) this gives
\begin{equation}
\label{vr1}
V(r) \sim -{\vec \sigma_1}\cdot{\vec \sigma_1} \ \alpha^{2-4z-4y} 
\int q^{2(z+y-1)} e^{i{\bf q r}}d^2 q 
\sim -{{{\vec \sigma_1}\cdot{\vec \sigma_1}}\over{r}}.
\end{equation}
This is a long range ferromagnetic interaction.
An interesting fact is that the interaction is independent of the bare coupling
constant $\alpha$. Another interesting fact is that there is no renormalization of the 
power in the $1/r$ dependence. However we stress that these facts have been proven only in 
the one loop approximation (one loop above the leading noncrossing 
approximation).

\section{conclusion}
The dynamics of the magnetic impurity in a 2D antiferromagnet close to
the critical point is highly nontrivial. To some extend it is similar to
the dynamics in the Kondo problem.
In the present paper we have considered spin 1/2 impurity.
We have calculated indexes for critical behavior of the Green's function,
vertexes, magnetic moments and magnetic susceptibilities.
We have also considered the impurity specific 
heat and  long range interaction between two impurities which is due to the
spin-wave exchange.

\acknowledgments
I am grateful to A.V. Chubukov, K. Le Hur, M. Yu. Kuchiev,  M. Troyer and 
especially to A. Sandvik for very helpful discussions.
The main part of this work has been done during my stay at ITP UCSB and ITP ETH Zurich.
It was supported by NSF Grant PHY94-07194 and by the Swiss National Fund.


\begin{references}
\bibitem{Lee} For a review see P. A. Lee, T. M. Rice, J. W. Serene, L. J. Sham,
and J. W. Wilkins, Comments Condense Matter Phys. {\bf 12}, 99 (1986).
\bibitem{dot} See, e. g., {\it Correlated Fermions and Transport in
Mesoscopic Systems}, edited by T. Martin, G. Montambaux, and J. Tran Tanh Van
(Editions Frontiers, Gif-sur-Yvette, France, 1996).
\bibitem{Zhu} See, e. g., Zhu-Pei Shi, R. R. P. Singh, M. P. Gelfand, 
and Ziqiang Wang, Phys. Rev. B {\bf 51}, 15630 (1995).
\bibitem{Nagaosa} N. Nagaosa, Y. Hatsugai, and M. Imada, J. Phys. Soc. Jap.
{\bf 58}, 978 (1989).
\bibitem{Igarashi} J. Igarashi, K. Murayama, and P. Fulde, Phys. Rev. B
{\bf 52}, 15966 (1995); K. Murayama and J. Igarashi, J. Phys. Soc. Jap.
{\bf 66}, 1157 (1997).
\bibitem{Oitmaa1} J. Oitmaa, D. Betts, and M. Aydin, Phys. Rev. B
{\bf 51}, 2896 (1995).
\bibitem{KLee} K. Lee and  P. Schlottmann, Phys. Rev. B
{\bf 42}, 4426 (1990).
\bibitem{Kotov1} V. N. Kotov, J. Oitmaa, O. P.  Sushkov, Phys. Rev. B
{\bf 58}, 8495 (1998); {\bf 58}, 8500 (1998).
\bibitem{hole} O. P. Sushkov, cond-mat/9808302.
\bibitem{Vojta} M. Vojta, C. Buragohain, and S. Sachdev, cond-mat/9912020.
\bibitem{note} We would like to note that there is another system with 
dynamics somewhat similar to that considred
in the present work. This is a nonmagnetic impurity in the vicinity of
O(1) quantum phase transition between plaquette and spin-dimerized phase,
see V. N. Kotov, M. E. Zhitomirsky, and O. P. Sushkov, cond-mat/0001282.
\bibitem{Hida} K. Hida, J. Phys. Soc. Jpn. {\bf 59}, 2230 (1990).
\bibitem{Sandvik} A. W. Sandvik and D. J. Scalapino, Phys. Rev. Lett. {\bf 72}, 
2777 (1994).
\bibitem{Jaklic} J. Jaklic and P. Prelovsek, Phys. Rev. Lett. {\bf 77}, 
892 (1996).
\bibitem{Weihong} Z. Weihong, Phys. Rev. B {\bf 55}, 12267 (1997).
\bibitem{Morr} D. Morr, A. V. Chubukov,
 Phys. Rev. B {\bf 52}, 3521 (1995).
\bibitem{Kotov} V. N. Kotov, O. Sushkov, Z. Weihong and J. Oitmaa, Phys. Rev. Lett. {\bf 80}, 5790 (1998)
\bibitem{Oitmaa} J. Oitmaa, R. R. P. Singh, and W. H. Zheng, Phys. Rev. B
{\bf 54}, 1009 (1996).
\bibitem{Elstner} N. Elstner and R. R. P. Singh, Phys. Rev. B
{\bf 57}, 7740 (1998).
\bibitem{SSS}P. V. Shevchenko, A. W. Sandvik, and O. P. Sushkov, 
Phys. Rev. B {\bf 61}, 3475 (2000).
\bibitem{BR} A. V. Chubukov, JETP Lett. {\bf 49}, 129 (1989);
S. Sachdev and R. Bhatt, Phys. Rev. B {\bf 41}, 9323 (1990). 
\bibitem{CSY} A. V. Chubukov, S. Sachdev, and J. Ye, Phys. Rev. B {\bf 49}, 11919 (1994).
\bibitem{Mats} To avoid misunderstanding we stress that we use finite temperature
  perturbation technique based on retarded Green's functions \cite{SSS}.
  We do not use Matsubara technique which is not convenient in this case.


\end{references}
\end{document}